\documentclass[%
 reprint,
superscriptaddress,
 amsmath,amssymb,
 aps,
prb,
showkeys,
]{revtex4-2}
\pdfoutput=1
\usepackage{graphicx}
\usepackage{dcolumn}
\usepackage{bm}
\usepackage[colorlinks=true, letterpaper=ture, pdfstartview=FitV, linkcolor=blue, citecolor=blue, urlcolor=blue]{hyperref}

\begin{document}

\preprint{APS/123-QED}

\title{Effect of physical and chemical pressure on the superconductivity of caged-type quasiskutterudite \texorpdfstring{Lu$_{5}$Rh$_{6}$Sn$_{18}$}{}}

\author{Chao Xiong}\thanks{These authors contributed to this work equally.}
 \affiliation{School of Physical Science and Technology, ShanghaiTech University, Shanghai 201210, China}
\author{Cuiying Pei}\thanks{These authors contributed to this work equally.}
 \affiliation{School of Physical Science and Technology, ShanghaiTech University, Shanghai 201210, China}
\author{Qi Wang}
 \affiliation{School of Physical Science and Technology, ShanghaiTech University, Shanghai 201210, China}
  \affiliation{ShanghaiTech Laboratory for Topological Physics, ShanghaiTech University, Shanghai 201210, China}
\author{Yi Zhao}
 \affiliation{School of Physical Science and Technology, ShanghaiTech University, Shanghai 201210, China}
\author{Yuyang Jiang}
 \affiliation{School of Physical Science and Technology, ShanghaiTech University, Shanghai 201210, China}
\author{Changhua Li}
 \affiliation{School of Physical Science and Technology, ShanghaiTech University, Shanghai 201210, China}
\author{Weizheng Cao}
 \affiliation{School of Physical Science and Technology, ShanghaiTech University, Shanghai 201210, China}
\author{Na Yu}
 \affiliation{School of Physical Science and Technology, ShanghaiTech University, Shanghai 201210, China}
\author{Lili Zhang}
 \affiliation{Shanghai Synchrotron Radiation Facility, Shanghai Advanced Research Institute, Chinese Academy of Sciences, Shanghai 201203, China}
\author{Yanpeng Qi}
 \email{qiyp@shanghaitech.edu.cn}
 \affiliation{School of Physical Science and Technology, ShanghaiTech University, Shanghai 201210, China}
 \affiliation{ShanghaiTech Laboratory for Topological Physics, ShanghaiTech University, Shanghai 201210, China}
 \affiliation{Shanghai Key Laboratory of High-Resolution Electron Microscopy, ShanghaiTech University, Shanghai 201210, China}

%

%

\date{\today}

\begin{abstract}
\begin{description}
\item[Abstract]
Lu$_{5}$Rh$_{6}$Sn$_{18}$ is one of the caged-type quasiskutterudite superconductors with superconducting transition temperature \textit{T}$_{c}$ = 4.12 K. Here, we investigate the effect of pressure on the superconductivity in Lu$_{5}$Rh$_{6}$Sn$_{18}$ by combining high pressure electrical transport, synchrotron x-ray diffraction (XRD) and chemical doping. Application of high pressure can enhance both the metallicity and the superconducting transition temperature in Lu$_{5}$Rh$_{6}$Sn$_{18}$. \textit{T}$_{c}$ is found to show a continuous increase reaching up to 5.50 K at 11.4 GPa. Our high pressure synchrotron XRD measurements demonstrate the stability of the pristine crystal structure up to 12.0 GPa. In contrast, \textit{T}$_{c}$ is suppressed after the substitution of La ions in Lu sites, inducing negative chemical pressure. Our study provides valuable insights into the improvement of superconductivity in caged compounds.
\end{description}
\end{abstract}

\maketitle


\section{Introduction}

Caged-type compounds exhibit three-dimensional structures characterized by the presence of spacious atomic cages enclosing comparatively smaller atoms\cite{67,AOs2O6,zhaoyi}. Because of strong electron-phonon coupling and weak structural coupling, the small atoms can `rattle' with large atomic excursions, ultimately leading to a rattling vibration. Three caged-type compounds, namely Si/Ge clathrates\cite{Si}, filled skutterudites (RT$_{4}$X$_{12}$)\cite{PrPt4Ge12,ReRu4Sb12,1412}, and $\beta$-pyrochlore oxides\cite{KOs2O6,CsOs2O6}, have been extensively investigated as ``rattling-good" materials. These compounds are renowned for a rich variety of properties\cite{fermion,transition,Orders,ReRu4Sb12,Pei-IrX3,BaxIr4X12}, such as heavy-fermion behavior, metal-insulator transition, multipole ordering, and superconductivity.

Ternary stannides R$_{5}$M$_{6}$Sn$_{18}$ (R = Rare earths, M = Transition metals)\cite{1,5,growth1980} can also be categorized as caged-type compounds, and some of them exhibit superconductivity (SC) with the transition temperature \textit{T}$_{c}$ = 4.4 K (Sc)\cite{21}, 4.0 K (Lu)\cite{15}, and 3.0 K (Y)\cite{11} when M = Rh. For the normal state, the resistivity \textit{$\rho$}(\textit{T}) of the Lu and Y compounds did not show a typical metallic behavior and exhibited an unusual temperature variation\cite{15,11}. Heat capacity measurements showed that Sc$_{5}$Rh$_{6}$Sn$_{18}$ and Lu$_{5}$Rh$_{6}$Sn$_{18}$ have an isotropic superconducting gap\cite{11,47}, while Y$_{5}$Rh$_{6}$Sn$_{18}$ has an anisotropic gap with point nodes\cite{11}. Interestingly, the muon spin relaxation ($\mu$SR) experiments indicated that both Lu$_{5}$Rh$_{6}$Sn$_{18}$ and Y$_{5}$Rh$_{6}$Sn$_{18}$ exhibit time-reversal symmetry (TRS) breaking in their superconducting state\cite{15,22-growth}. In addition, coexistence of superconductivity and magnetism was observed in the reentrant superconductors Tm$_{5}$Rh$_{6}$Sn$_{18}$ (\textit{T}$_{c}$ = 2.20 K)\cite{52} and Er$_{5}$Rh$_{6}$Sn$_{18}$ (\textit{T}$_{c}$ = 1.05 K)\cite{10} with the same crystal structure.

Given the presence of covalently bonded cage-forming frameworks and the rattling motion of the guest atoms, the caged compounds are in principle sensitive to the external conditions. The application of pressure to caged compounds has led to a profound effect on their physical properties, e.g. superconductivity\cite{pei,BaIrGe}. \textit{T}$_{c}$ decreases with increasing pressure for most skutterudite related superconductors, showing a negative pressure coefficient of \textit{T}$_{c}$\cite{19}. Particularly, the positive pressure coefficient observed in La$_{3}$Ru$_{4}$Sn$_{13}$\cite{LaRuSn} and Sc$_{5}$Rh$_{6}$Sn$_{18}$\cite{29} with the rate of 0.03 K/GPa and 0.10 K/GPa, respectively. It should be noted that those high pressure studies performed at very limited pressure range, it is intriguing to explore whether this improvement can be sustained at higher pressures.

Motivated by the above issues, we systematically investigate the effect of pressure on skutterudite-related superconductor. We chose Lu$_{5}$Rh$_{6}$Sn$_{18}$ as a model system and studied the evolution of superconductivity when applying physical pressure. In order to compare the results, we also conducted chemical doping and discussed the effect of negative chemical pressure on the Lu$_{5}$Rh$_{6}$Sn$_{18}$ superconductor.

\section{Experimental details}

Single crystals of La$_{x}$Lu$_{5-x}$Rh$_{6}$Sn$_{18}$(x = 0, 0.5 and 1) were grown through a normal Sn-flux method\cite{growth1980,growth,22-growth}. High-purity Lu shot (99.9\%), La shot (99.9\%), Rh powder (99.9\%), and Sn shot (99.999\%) were mixed in the proportion of Lu:La:Rh:Sn = 5-x:x:10:100. The mixture was then loaded into an alumina crucible and then stored in a quartz tube at a glove box under argon atmosphere. The materials were melted together to 1050 $^{\circ}$C for about 24 hours, held at 1050 $^{\circ}$C for about 3 hours, and cooled down to 575 $^{\circ}$C at a speed of 5 $^{\circ}$C/h. The La concentration is limited up to x = 1 since impurity phases become apparent with increasing La concentration. The single crystals La$_{x}$Lu$_{5-x}$Rh$_{6}$Sn$_{18}$(x = 0, 0.5 and 1) have a shiny surface with typical size of 3 $\times$ 2 $\times$ 2 mm$^{3}$.

Phase and quality examinations of the single crystals were performed on Bruker D8 single-crystal x-ray diffractometer (Mo\textit{K}$_{\alpha}$, $\lambda$ = 0.71073 {\AA}) and Bruker D2 powder x-ray diffractometer (Cu\textit{K}$_{\alpha}$, $\lambda$ = 1.54056 {\AA}). Rietveld refinements of the powder XRD pattern were analyzed utilizing the General Structure Analysis System (GSAS-II)\cite{GSAS}. The composition and microstructure of the samples were characterized by a scanning electron microscope (Phenom Pro) equipped with energy dispersive spectroscopy (EDS).

Electrical resistivity and magnetization measurements from 1.8 K to 300 K were achieved using PPMS and MPMS (Quantum Design), respectively. A standard dc four-probe method with Au wires was adopted to gain the electrical resistivity at ambient pressure. The dc magnetization was measured after both zero-field cooling (ZFC) and field cooling (FC) in an applied magnetic field of 10 Oe and the ac magnetic susceptibility was measured at 1000 Hz and with amplitude of 2 Oe at various temperatures.

High pressure resistivity measurements were performed in a BeCu-type diamond anvil cell (DAC) with 400 $\mu$m culet as described elsewhere\cite{pei,wangqi}. A cubic BN/epoxy mixture layer was inserted between the BeCu gasket and the electrical leads. Four platinum foils were placed with the sample following the van der Pauw method\cite{WB2,MoB2}. High pressure XRD experiments were performed at beamline BL15U of the Shanghai Synchrotron Radiation Facility ($\lambda$ = 0.6199 \r{A}). Mineral oil was used as a pressure transmitting medium. Rietveld refinement was accomplished employing the GSAS-II to determine the structural parameters. Pressure was measured using the ruby luminescence method\cite{rubby}.

\section{Results}

\begin{figure}[!tbp]
\centering
\includegraphics[width=8.6cm]{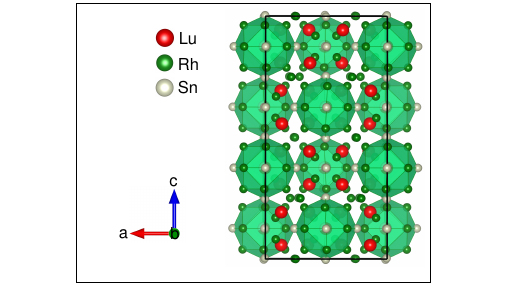}
\caption{\label{fig_1} The crystal structure of Lu$_{5}$Rh$_{6}$Sn$_{18}$. The occupation parameter of Lu1 is determined to be 0.69.}
\end{figure}

\begin{table}[!tbp]
\caption{\label{tab:table1}%
Crystallographic data of Lu$_ {5}$Rh$_{6}$Sn$_{18}$.}
\begin{ruledtabular}
\begin{tabular}{ll}
Formula & Lu$_ {4.69}$Rh$_{6}$Sn$_{18}$\\
\hline
Space group & \textit{I}4$_{1}$/\textit{acd} (no. 142) \\
Temperature [K] & 278.00 \\
Formula per unit cell, Z & 8 \\
2$\theta$ range for data collection/$^{\circ}$ & 5.158 to 52.722 \\
\textit{a}, \textit{b} [\r{A}] & 13.6829(3)\\
\textit{c} [\r{A}] &  27.3503(9)\\
Unit cell volume, V [{\AA}$^{3}$] & 5120.6(3) \\
Calculated density, $\rho$(g/cm$^{3}$) & 9.273\\
$\mu$/mm$^{-1}$ &38.797\\
F(000) & 12023.0\\
Collected reflections &1313 \\
Independent reflections &1313 \\
Goodness-of-fit on F$^{2}$ &1.046\\
\textit{R}$_{1}$/\textit{wR}$_{2}$, I $\geq$ 2$\sigma$(I) & 0.0422/0.1143 \\
\textit{R}$_{1}$/\textit{wR}$_{2}$ [all data] & 0.0529/0.1217 \\
Largest diff. peak/hole/(e \r{A}$^{-3}$) &2.688/-3.54 \\
\end{tabular}
\end{ruledtabular}
\end{table}

\begin{figure}[!tbp]
\centering
\includegraphics[width=8.6cm]{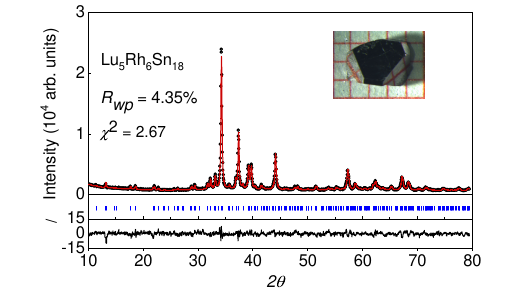}
\caption{\label{fig_2} Powder XRD pattern of Lu$_{5}$Rh$_{6}$Sn$_{18}$. The solid lines represent the Rietveld fits for Lu$_{5}$Rh$_{6}$Sn$_{18}$, while the open circles correspond to the observed data. The solid lines at the bottom indicate the residual intensities. The inset shows photo of the Lu$_ {5}$Rh$_{6}$Sn$_{18}$ single crystals. }
\end{figure}

It has been reported that the Lu$_{5}$Rh$_{6}$Sn$_{18}$ single crystals exhibit a high degree of disorder and twinning\cite{18}. In order to address this issue, single crystal XRD measurements are performed at room temperature. Using the twinning tool of the program OLEX2\cite{OLEX2}, the structure was solved with the SHELXT\cite{SHELXT} structure solution program using Intrinsic Phasing and refined with the SHELXL\cite{SHELXL} refinement package using Least Squares minimisation. The details of the structural refinements are displayed in Table~\ref{tab:table1} and \ref{tab:table2}. The tetragonal structure of Lu$_{5}$Rh$_{6}$Sn$_{18}$ is depicted in Fig. \ref{fig_1}. To check the result analyzed based on single crystal XRD, powder XRD measurements were performed on the same batch of Lu$_{5}$Rh$_{6}$Sn$_{18}$. Fig. \ref{fig_2} shows the refined powder XRD pattern of Lu$_{5}$Rh$_{6}$Sn$_{18}$. The Bragg reflections can be well refined using the obtained structure with reliable parameters. The consistence between powder and single crystal XRD measurements guarantees the correct phase. The crystal structure of Lu$_{5}$Rh$_{6}$Sn$_{18}$ exhibits a distortion of a cubic structure (space group \textit{Fm-}3\textit{m}), resulting in an approximate doubling of the length of the \textit{c}-axis\cite{17}. Lu1 atoms are located inside cuboctahedras and surrounded by the Rh$_{6}$Sn$_{18}$ cage. The refined results indicate the existence of vacancies at Lu sites (0.69) and agree well with the EDS measurements (Fig. S1 in the Supplemental Material\cite{supp}). No trace of superstructure peaks can be observed from the powder XRD pattern, which means the Lu vacancies tend to be randomly distributed in the lattice. Our data, in agreement with neutron diffraction measurements\cite{18,56}, illustrate the highly twinned nature of the crystals, with twin domain scales of 0.53. The disorder of Lu$_{5}$Rh$_{6}$Sn$_{18}$ may be attributed to a micro-twinned stacking disorder of (A, B) square planes\cite{10,17}.

\begin{figure*}[!tbp]
\centering
\includegraphics{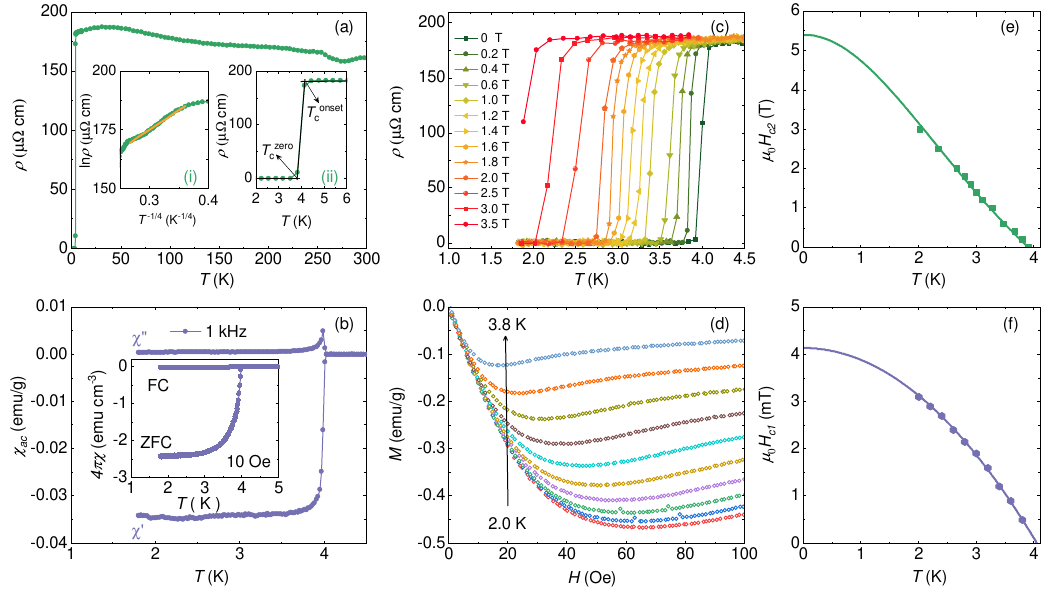}
\caption{\label{fig_3}(a) Resistivity of Lu$_ {5}$Rh$_{6}$Sn$_{18}$ in zero field from 1.8 K to 300 K. The inset (i) displays the resistivity data in coordinates ln$\rho$ = \textit{f} (\textit{T}$^{-1/4}$). Yellow line approximates the linear behavior in the temperature range between 60 K and 200 K. The inset (ii) illustrates the superconducting transitions at low temperatures. (b) Temperature dependencies of the real and imaginary part of the ac magnetic susceptibility of Lu$_{5}$Rh$_{6}$Sn$_{18}$. The inset shows the dc magnetic susceptibility in ZFC and FC conditions. (c) Electrical resistivity at various magnetic field for Lu$_ {5}$Rh$_{6}$Sn$_{18}$. (d) Field dependence of the magnetization \textit{M}(\textit{H}) at different temperatures below \textit{T}$_{c}$. (e) and (f) Temperature dependence of \textit{$\mu_{0}$H}$_{c2}$ and \textit{$\mu_{0}$H}$_{c1}$ for Lu$_ {5}$Rh$_{6}$Sn$_{18}$.}
\end{figure*}

Fig. \ref{fig_3}(a) shows the temperature dependence of the electrical resistivity \textit{$\rho$}(\textit{T}) of Lu$_{5}$Rh$_{6}$Sn$_{18}$. The normal state of Lu$_{5}$Rh$_{6}$Sn$_{18}$ exhibits unusual temperature variation: \textit{$\rho$}(\textit{T}) is nearly independent of \textit{T} down to about 150 K, and shows an increase on further cooling. The negative temperature coefficient of the resistivity (d$\rho$/d$\textit{T}$ $<$ 0) over a wide temperature range with ln\textit{$\rho$} $\sim$ \textit{T}$^{-1/4}$ dependence (inset (i) of Fig. \ref{fig_3}(a)) is in line with previous reports\cite{26}, which is known as the Mott variable-range hopping effect\cite{Mott}. Recently theoretical calculations for Lu$_{5}$Rh$_{6}$Sn$_{18}$ documented the pseudogap with very small density of states (DOS)\cite{18}, located in the bands of these compounds at a similar binding energy of $\sim$ -0.3 eV. Even a small number of vacancies shifts this pseudogap toward the Fermi level\cite{26}. Our single crystal XRD measurements demonstrate the existence of Lu vacancies. Therefore, the negative temperature coefficient effect in \textit{$\rho$}(\textit{T}) for Lu$_{5}$Rh$_{6}$Sn$_{18}$ single crystals results from the off-stoichiometry effect.

A sharp superconduting transition occurs at low temperature with a residual resistivity \textit{$\rho$}$_{0}$ around 181 $\mu\Omega$cm obtained from extreapolating the normal state to zero temperature. The transition temperatures \textit{T}$^{onset}_{c}$ and \textit{T}$^{zero}_{c}$ of Lu$_{5}$Rh$_{6}$Sn$_{18}$ are 4.12 K and 3.90 K (inset (ii) of Fig. \ref{fig_3}(a)), respectively. Fig. \ref{fig_3}(b) displays the temperature dependencies of the real (\textit{$\chi^{'}$}) and imaginary (\textit{$\chi^{''}$}) parts of the ac magnetic susceptibility \textit{$\chi_{ac}$} of Lu$_{5}$Rh$_{6}$Sn$_{18}$. The inset shows the dc magnetic susceptibility of Lu$_ {5}$Rh$_{6}$Sn$_{18}$ crystal measured with magnetic field \textit{H} = 10 Oe in ZFC and FC conditions. A clear onset of the diamagnetic shift was observed at \textit{T}$^{mag}_{c}$ = 4.05 K, in agreement with the resistivity result. The large diamagnetic signal at 1.8 K confirms the bulk superconductivity. Fig. \ref{fig_3}(c) shows the suppression of \textit{T}$_{c}$ with the increase of the external magnetic field. The upper critical field, $\mu_{0}$\textit{H}$_{c2}$, is determined using the point where \textit{$\rho$} = 0 on the resistivity transition curve, and plots of $\mu_{0}$\textit{H}$_{c2}$(\textit{T}) are depicted in Fig. \ref{fig_3}(e). A simple estimation of the \textit{$\mu_{0}$}\textit{H}$_{c2}$(0) from the Ginzburg-Landau (GL) formula\cite{HC2-1}
\begin{equation}\label{1}
\mu_{0}\textit{H}_{c2}(T) =\mu_{0}\textit{H}_{c2}(0)\frac{1 - t^{2}}{1 + t^{2}},
\end{equation}
t = \textit{T}/\textit{T}$_{c}$, which gave an upper critical field of 5.40 T. The coherence length $\xi$ can be acquired from Ginzburg-Landau (GL) equation
\begin{equation}\label{4}
\mu_{0}\textit{H}_{c2} = \frac{\Phi }{2\pi\xi^{2}},
\end{equation}
where $\Phi_{0}$ is the magnetic flux unit. By using $\mu_{0}$\textit{H}$_{c2}$(0) = 5.40 T, the calculated $\xi$ should be 7.81 nm.
The magnetization versus external field over a range of temperatures below \textit{T}$_{c}$ is presented in Fig. \ref{fig_3}(d). The field deviating from a linear curve of full Meissner effect was deemed as the lower critical field $\mu_{0}$\textit{H}$_{c1}$ at each temperature and was summarized in Fig. \ref{fig_3}(f). $\mu_{0}$\textit{H}$_{c1}$(0) is found to be 4.14 mT by using the Ginzburg-Landau (GL) formula
\begin{equation}\label{2}
\mu_{0}\textit{H}_{c1}(T) = \mu_{0}\textit{H}_{c1}(0)\left[1 - \left(\frac{T}{T_{c}}\right)^{2}\right].
\end{equation}
By using formula
\begin{equation}\label{3}
\mu_{0}\textit{H}_{c2} = \frac{\Phi}{4\pi\lambda^{2}}ln\left(\frac{\lambda}{\xi}\right),
\end{equation}
we get the penetration depth $\lambda$ = 395 nm. The calculated GL parameter of $\kappa$ = $\lambda$/$\xi$ $\sim$ 50.58 confirms the type-II superconductivity in Lu$_ {5}$Rh$_{6}$Sn$_{18}$. The thermodynamic critical fields \textit{$\mu_{0}$}\textit{H}$_{c}$(0) can be obtained by $\mu_{0}$\textit{H}$_{c2}$(0)/$\sqrt{2}\kappa$ and all the physical parameters are summarized in Table~\ref{tab:table3}.

\begin{table}[!tbp]
\caption{\label{tab:table2}%
Atomic coordinates, occupations of Lu$_ {5}$Rh$_{6}$Sn$_{18}$ from single crystal XRD data.}
\begin{ruledtabular}
\begin{tabular}{cccccccccccc}
Atom & Site & Occ. & x & y & z & \\
\hline
Lu1 & 8b & 0.69 & 0.50000 & 0.75000 & 0.62500 \\
Lu2 & 32g & 1.0 & 0.63268 & 0.88658 & 0.80712 \\
Rh1 & 16d & 1.0 & 0.50000 & 0.75000 & 0.75313 \\
Rh2 & 32g & 1.0 & 0.50095 & 0.49344 & 0.62493 \\
Sn1 & 32g & 1.0 &  0.49480 & 0.57402 & 0.53797 \\
Sn2 & 32g & 1.0 & 0.50919 & 0.88658 & 0.80712 \\
Sn3 & 16f & 1.0 & 0.32623 & 0.57623 & 0.62500 \\
Sn4 & 16f & 1.0 & 0.67637 & 0.57363 & 0.62500 \\
Sn5 & 32g & 1.0 & 0.41254 & 0.33864 & 0.58090 \\
Sn6 & 16e & 1.0 & 0.78794 & 1.00000 & 0.75000 \\
\end{tabular}
\end{ruledtabular}
\end{table}

\begin{figure*}[!tbp]
\centering
\includegraphics{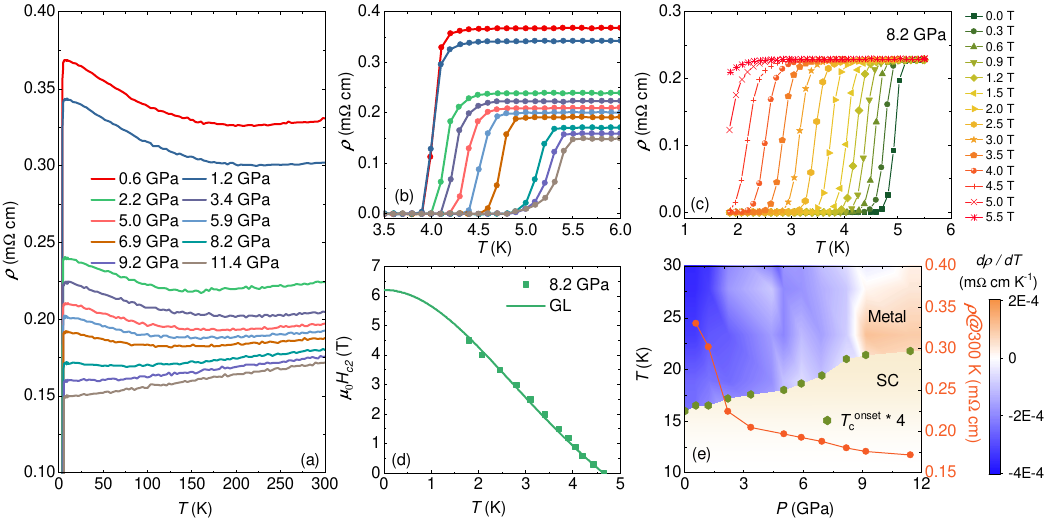}
\caption{\label{fig_4} Characterization of the superconducting transition of Lu$_{5}$Rh$_{6}$Sn$_{18}$ at various pressures. (a) Resistivity toward temperature obtained from the measurements in the pressure between 0.6 and 11.4 GPa, from 1.8 K to 300 K. (b) Detail of the resistivity at low temperatures for the data in (a), clearly indicating the impact of pressure on the resistivity through the superconducting transition and the preservation of the zero-resistance state.  (c) Magnetic field dependence of the superconducting transition in Lu$_{5}$Rh$_{6}$Sn$_{18}$ at 8.2 GPa. (d) \textit{$\mu_{0}$H}$_{c2}$ as a function of temperature. The solid line represent the Ginzburg–Landau fits. (e) Phase diagram of Lu$_{5}$Rh$_{6}$Sn$_{18}$. The red dots show the variation in resistivity with pressure at 300 K. Green hexagon represent superconducting $\textit{T}_{c}$. The background represents the dependence of d$\rho$/d\textit{T} from 10 K to 30 K at different pressures.}
\end{figure*}

\begin{figure}[!tbp]
\centering
\includegraphics[width=8.6cm]{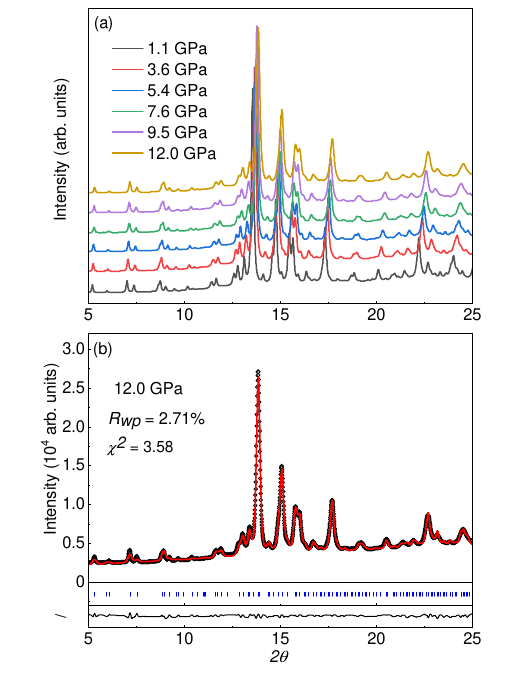}
\caption{\label{fig_5} (a) Powder XRD patterns of Lu$_{5}$Rh$_{6}$Sn$_{18}$ at various pressures up to 12.0 GPa. (b) XRD pattern together with the Rietveld refinement results at 12.0 GPa.}
\end{figure}

\begin{figure}[!tbp]
\centering
\includegraphics[width=8.6cm]{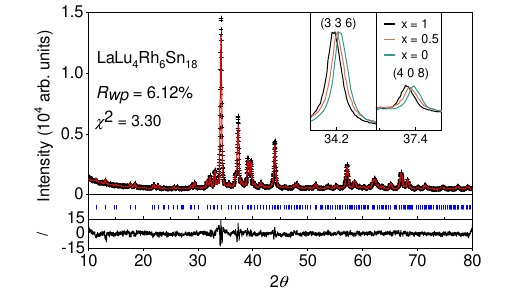}
\caption{\label{fig_6} Powder XRD pattern of LaLu$_{4}$Rh$_{6}$Sn$_{18}$. The insets show the appearance of the (3 3 6) and (4 0 8) reflection of the x = 0, 0.5 and 1 samples.}
\end{figure}

\begin{figure*}[!tbp]
\centering
\includegraphics{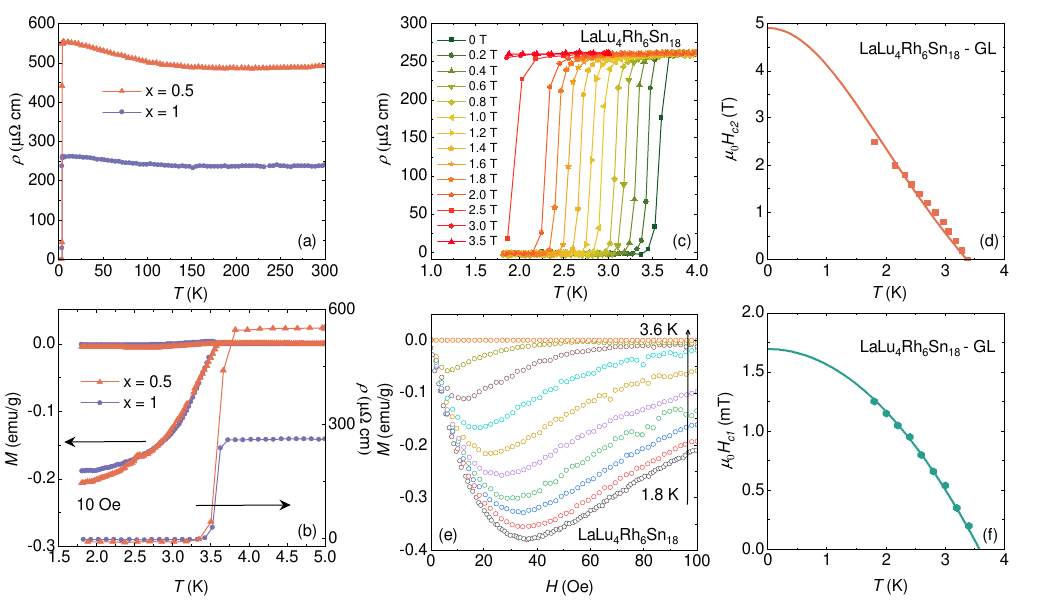}
\caption{\label{fig_7} (a) Resistivity of La$_ {x}$Lu$_ {5-x}$Rh$_{6}$Sn$_{18}$(x = 0.5 and 1) in zero field from 1.8 K to 300 K. (b) Superconducting transitions at low temperatures and the dc magnetic susceptibility in ZFC and FC conditions. (c) Electrical resistivity at various applied magnetic field for LaLu$_ {4}$Rh$_{6}$Sn$_{18}$. (d) Field dependence of the magnetization \textit{M}(\textit{H}) of LaLu$_ {4}$Rh$_{6}$Sn$_{18}$ at different temperatures below \textit{T}$_{c}$. (e) and (f) Temperature dependence of \textit{$\mu_{0}$H}$_{c2}$ and \textit{$\mu_{0}$H}$_{c1}$ for LaLu$_ {4}$Rh$_{6}$Sn$_{18}$.}
\end{figure*}

\begin{figure}[!tbp]
\centering
\includegraphics{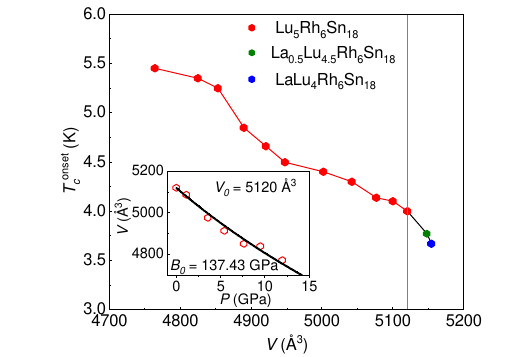}
\caption{\label{fig_8} \textit{T}$_{c}$ as a function of volume of La$_ {x}$Lu$_ {5-x}$Rh$_{6}$Sn$_{18}$(x = 0, 0.5, 1). Red points corresponds to \textit{T}$_{c}^{onset}$ in the context of our high pressure studies of Lu$_{5}$Rh$_{6}$Sn$_{18}$. The inset shows the unit cell volume of Lu$_{5}$Rh$_{6}$Sn$_{18}$ up to around 12.0 GPa as a function of pressure. Black line represents the fitting result.}
\end{figure}

\begin{table}[!tbp]
\caption{\label{tab:table3}%
Superconducting parameters of Lu$_ {5}$Rh$_{6}$Sn$_{18}$ and LaLu$_ {4}$Rh$_{6}$Sn$_{18}$.}
\begin{ruledtabular}
\begin{tabular}{ccccccc}
 & Lu$_ {5}$Rh$_{6}$Sn$_{18}$ & LaLu$_ {4}$Rh$_{6}$Sn$_{18}$ \\
\hline
\textit{T}$_{c}^{zero}$ (K) & 3.90 & 3.33 \\
\textit{T}$_{c}^{mag}$ (K) & 4.05 & 3.58 \\
\textit{T}$_{c}^{onset}$ (K) & 4.12 & 3.72 \\
\textit{$\mu_{0}$}\textit{H}$_{c2}$(0) (T) & 5.40 & 4.90 \\
\textit{$\mu_{0}$}\textit{H}$_{c1}$(0) (mT)  & 4.14 & 1.68 \\
\textit{$\mu_{0}$}\textit{H}$_{c}$(0) (mT) & 75.49 & 43.48 \\
$\xi$ (nm) & 7.81 & 8.20 \\
$\lambda$ (nm) & 395 & 655 \\
$\kappa$ & 50.58 & 79.88 \\
\end{tabular}
\end{ruledtabular}
\end{table}

\begin{table}[!tbp]
\caption{\label{tab:table4}%
Chemical composition, lattice constants and volume of La$_{x}$Lu$_ {5-x}$Rh$_{6}$Sn$_{18}$.}
\begin{ruledtabular}
\begin{tabular}{cccccccc}
x & Composition (La:Lu:Rh:Sn) & \textit{a}, \textit{c} (\r{A}) & \textit{V} \\
\hline
0 & 0.00 : 3.84 : 6.00 : 18.12  & 13.6829, 27.3503 & 5120 \\
0.5 & 0.83 : 2.62 : 6.00 : 16.68 & 13.7095, 27.3885 & 5148 \\
1 & 0.97 : 2.88 : 6.00 : 16.53 & 13.7142, 27.4039  & 5154\\
\end{tabular}
\end{ruledtabular}
\end{table}

In order to determine the pressure coefficient of $\textit{T}_{c}$ in Lu$_{5}$Rh$_{6}$Sn$_{18}$, we measured the electrical resistivity \textit{$\rho$}(\textit{T}) of Lu$_{5}$Rh$_{6}$Sn$_{18}$ at various pressures. Fig. \ref{fig_4}(a) shows the typical \textit{$\rho$}(\textit{T}) curves of Lu$_{5}$Rh$_{6}$Sn$_{18}$ for pressures up to 11.4 GPa. Increasing pressure induces a continuous suppression of the overall magnitude of \textit{$\rho$}. At lower temperature, the upturn of \textit{$\rho$}(\textit{T}) fades away gradually and the semimetallic behavior is being suppressed by the application of pressure. Above 8.2 GPa, the \textit{$\rho$}(\textit{T}) of Lu$_{5}$Rh$_{6}$Sn$_{18}$ exhibits typical metallic behavior. This is similar to the metallic behavior observed in Y$_{5}$Rh$_{6}$Sn$_{18}$ doped with La or Ti over the whole temperature range\cite{50}. The band structure calculations confirm the \textit{d}-band character of the conduction electrons, which dominate the DOS at Femi level\cite{18,25,45}. The enhancement of the metallicity is likely attributed to the pressure increased Coulomb interaction between the \textit{d} electrons of transition metal, which dominate the field-dependent electronic transport in these materials\cite{25}. Fig. \ref{fig_4}(b) shows the effect of pressure on the resistivity during the superconducting transition. \textit{T}$_{c}$ increases with pressure and a maximum \textit{T}$^{onset}_{c}$ of 5.50 K is attained at 11.4 GPa, revealing a positive pressure coefficient of $\textit{T}_{c}$. An overall increment rate of d\textit{T}$_{c}$/d$\textit{P}$ = 0.13 K GPa$^{-1}$ is shown in Lu$_{5}$Rh$_{6}$Sn$_{18}$ single crystals, which is comparable with other similar tetragonal skutterudites\cite{25,29}. Fig. \ref{fig_4}(c) demonstrates that the resistivity drop is continuously suppressed with increasing magnetic field. The value of $\mu_{0}$\textit{H}$_{c2}$(0) was estimated to be 6.20 T at 8.2 GPa (Fig. \ref{fig_4}(d)), which yields a coherence length $\xi$ of 7.29 nm. The pressure dependence of \textit{T}$_{c}$ for Lu$_{5}$Rh$_{6}$Sn$_{18}$ is summarized in Fig. \ref{fig_4}(e). Here, the resistivity values at 300 K are also shown. \textit{T}$_{c}$ increases with pressure and tends to saturate at pressure, where the temperature dependence of the resistivity changes from a semimetal behavior to that of a normal metal. The coincidence of the pressure-induced metallicity and enhancement of \textit{T}$_{c}$ suggests a competitive correlation of superconductivity and the semimetal state.

To further identify the structural stability of Lu$_{5}$Rh$_{6}$Sn$_{18}$, we performed \textit{in situ} high pressure XRD measurements. Fig. \ref{fig_5}(a) illustrates the high pressure synchrotron XRD patterns of Lu$_{5}$Rh$_{6}$Sn$_{18}$ measured at room temperature and pressures up to 12.0 GPa. These patterns have been adjusted vertically for better data presentation. A representative refinement obtained at 12.0 GPa is displayed in Fig. \ref{fig_5}(b). All the Bragg reflections can be refined by using the space group $\textit{I}4_{1}$\textit{/acd} as the initial model. As shown in Fig. S3 in the Supplemental Material\cite{supp}, both the \textit{a}-axis and \textit{c}-axis lattice parameters decrease with increasing pressure. A third-order Birch-Murnaghan equation of state\cite{Birch-Murnaghan} was used to fit the measured pressure-volume (\textit{P}-\textit{V}) data for Lu$_{5}$Rh$_{6}$Sn$_{18}$. The equation is given by:
\begin{eqnarray}
{P(V)}=&&\frac{3B_0}{2} \left[\left(\frac{V_0}{V}\right)^{7/3} - \left(\frac{V_0}{V}\right)^{5/3}\right]\nonumber\\
&&\times
\left\{1+\frac{3}{4}\left(B_0^\prime-4\right) \left[\left(\frac{V_0}{V}\right)^{2/3} - 1\right]\right\},
\end{eqnarray}
where \textit{B}$_{0}$ is the bulk modulus at ambient pressure, \textit{B}$_{0}^\prime$ is the pressure derivative of \textit{B}$_{0}$, \textit{V}$_{0}$ is the volume at ambient pressure. The obtained bulk modulus \textit{B}$_{0}$ is 137.43 GPa with \textit{V}$_{0}$ = 5120 \r{A}$^{3}$. Our synchrotron diffraction indicates that the structure of Lu$_{5}$Rh$_{6}$Sn$_{18}$ is robust until 12.0 GPa and rule out the possibility of any structural phase transition. This finding indicates that the enhanced superconductivity under pressure originates from the pristine tetragonal phase.

Our high pressure transport measurements demonstrate that lattice shrinkage is beneficial for improving superconductivity in Lu$_{5}$Rh$_{6}$Sn$_{18}$. To verify this scenario, we induce negative pressure by substituting the larger ion on the Lu site. Since La atoms have a relatively larger ionic radius, we chose La as the chemical dopant and grew a series of La$_{x}$Lu$_{5-x}$Rh$_{6}$Sn$_{18}$ (x = 0.5 and 1) single crystals. From the EDS analysis (Fig. S1 in the Supplemental Material\cite{supp}), we indeed found the doping elements, i.e. La, in our single crystals. The powder XRD patterns of La$_{x}$Lu$_{1-x}$Rh$_{6}$Sn$_{18}$ (x = 0.5 and 1) (Fig. S2 and Fig. \ref{fig_6}) show that all of the peaks can be well refined with the \textit{I}4$_{1}$/\textit{acd} space group without any impurity phases. As shown in the insets of Fig. \ref{fig_6}, the typical peaks of La-doped samples shift to lower angles, suggesting a lattice expansion. Details of the lattice constants are given in Table~\ref{tab:table4}. The evolution of crystal lattices together with the EDS results demonstrates the successful chemical substitution in La$_{x}$Lu$_{5-x}$Rh$_{6}$Sn$_{18}$.

Fig. \ref{fig_7}(a) shows the resistivity \textit{$\rho$}(\textit{T}) of the La$_{x}$Lu$_{5-x}$Rh$_{6}$Sn$_{18}$ (x = 0.5 and 1) single crystals. The doping of La does not enhance the metallicity, meanwhile, the superconducting transition temperature is suppressed to \textit{T}$^{onset}_{c}$ = 3.83 K and 3.72 K for x = 0.5 and 1, respectively. The temperature dependence of the magnetic susceptibility is shown in Fig. \ref{fig_7}(b) for both ZFC and FC conditions in an applied magnetic field of 10 Oe. The sharp superconducting transition is observed in both samples with the onset of diamagnetism occurring at \textit{T}$^{mag}_{c}$ = 3.63 K and 3.58 K for x = 0.5 and 1, respectively. We also performed the field dependence of \textit{$\rho$}(\textit{T}) and \textit{M}(\textit{H}) measurements for LaLu$_{4}$Rh$_{6}$Sn$_{18}$ as shown in Figs. \ref{fig_7}(c) and (d). The values of \textit{$\mu_{0}$}\textit{H}$_{c2}$(0) and \textit{$\mu_{0}$}\textit{H}$_{c1}$(0) are estimated to be 4.90 T and 1.68 mT, respectively. The details of superconducting parameters are also summarized in Table~\ref{tab:table3}.

\section{Discussions}

At last, we discuss the effect of pressure on the superconductivity in Lu$_{5}$Rh$_{6}$Sn$_{18}$. Fig. \ref{fig_8} plots the \textit{T}$^{onset}_{c}$ of La$_{x}$Lu$_{5-x}$Rh$_{6}$Sn$_{18}$(x = 0, 0.5, 1) as a function of the unit cell volume. It is clear to see the positive correlation between \textit{T}$_{c}$ and lattice volume in La$_{x}$Lu$_{5-x}$Rh$_{6}$Sn$_{18}$. Application of high pressure is found to enhance both the metallicity and the superconducting transition temperature. Recently, high pressure Raman spectrum  measurements in the isostructural compound Sc$_{5}$Rh$_{6}$Sn$_{18}$ have shown that there is no pressure-induced phonon softening linked to cage shrinkage, but a normal linear increase in the phonon mode frequencies\cite{29}. Therefore, the improvement of \textit{T}$_{c}$ in Lu$_{5}$Rh$_{6}$Sn$_{18}$ under pressure can be attributed to an enhancement in DOS at the Fermi level. On the other hand, the introduction of La into Lu$_{5}$Rh$_{6}$Sn$_{18}$ induced a negative chemical pressure, which reasonably plays the opposite effect. In addition, La is larger than Lu, the substitution of La ions would alter the rattling motion of the R atoms within the three-dimensional skeleton, which is detrimental to superconductivity. It should be mentioned that the suppression of superconductivity in La-doped Lu$_{5}$Rh$_{6}$Sn$_{18}$ is contrary to the behavior observed in isostructural compounds\cite{50,25,26}, where local atomic disorder leads to an abnormal increase in the superconducting transition temperature \textit{T}$_{c}$. Systematic studies from both experimental and theoretical perspectives are needed to understand the evolution of superconductivity in quasiskutterudites with various rare-earth atoms.

\section{Conclusions}

In summary, we systematically investigate the evolution of superconductivity in Lu$_{5}$Rh$_{6}$Sn$_{18}$ by applying physical pressure and chemical pressure. Application of pressure effectively improves both the metallic state and the superconductivity in Lu$_{5}$Rh$_{6}$Sn$_{18}$. The high pressure XRD data revealed that the cage structure was stable without any structural phase transition up to $\sim$ 12.0 GPa. On the other hand, $\textit{T}_{c}$ was suppressed when negative chemical pressure was induced. Our results indicate the positive correlation between \textit{T}$_{c}$ and lattice volume in La$_{x}$Lu$_{5-x}$Rh$_{6}$Sn$_{18}$, which will provide critical insight in similar caged superconductors.

\begin{acknowledgments}
This work was supported by the National Natural Science Foundation of China (Grant Nos. 52272265, U1932217, 11974246, 12004252), the National Key R$\&$D Program of China (Grant No. 2018YFA0704300), and Shanghai Science and Technology Plan (Grant No. 21DZ2260400). The authors thank the support from Analytical Instrumentation Center ($\#$ SPSTAIC10112914), SPST, ShanghaiTech University. The authors thank the staff members at BL15U1 in Shanghai Synchrotron Radiation Facility for assistance during data collection.
\end{acknowledgments}

\appendix

\bibliography{manuscript}

\end{document}